\documentclass[apjl]{emulateapj}
\usepackage{amsmath}
\usepackage{color}
\usepackage{ulem}

\slugcomment{ApJ Letter (accepted)}
\shorttitle{Unusual Optical Transient}
\shortauthors{Urata, Tsai et al.}

\begin{document}

\title{Unusual Long and Luminous Optical Transient in the Subaru Deep Field}

\author{
Yuji~\textsc{Urata}\altaffilmark{1}, 
Patrick~P.\textsc{Tsai}\altaffilmark{1}, 
Kuiyun~\textsc{Huang}\altaffilmark{2},
Tomoki~\textsc{Morokuma}\altaffilmark{3},
Naoki~\textsc{Yasuda}\altaffilmark{4}
Masaomi~\textsc{Tanaka}\altaffilmark{5},
Kentaro~\textsc{Motohara}\altaffilmark{3},
Masao~\textsc{Hayashi}\altaffilmark{5},
Nobunari~\textsc{Kashikawa}\altaffilmark{5},
Chun~\textsc{Ly}\altaffilmark{6,7}
and 
Matthew~A.\textsc{Malkan}\altaffilmark{8}
}

\altaffiltext{1}{Institute of Astronomy, National Central University, Chung-Li 32054, Taiwan, urata@astro.ncu.edu.tw}
\altaffiltext{2}{Academia Sinica Institute of Astronomy and Astrophysics, Taipei 106, Taiwan}
\altaffiltext{3}{Institute of Astronomy, Graduate School of Science, University of Tokyo, Mitaka, Tokyo 181-0015, Japan}
\altaffiltext{4}{Institute for the Physics and Mathematics of the Universe, University of Tokyo, Kashiwa 277-8568, Japan}
\altaffiltext{5}{Optical and Infrared Astronomy Division, National Astronomical Observatory, Mitaka, Tokyo 181-8588, Japan}
\altaffiltext{6}{Space Telescope Science Institute, Baltimore, MD, USA}
\altaffiltext{7}{Giacconi Fellow.}
\altaffiltext{8}{Department of Physics and Astronomy, UCLA, Box 951547, Los Angeles, CA, USA}

\begin{abstract}

  We present observations of SDF-05M05, an unusual optical transient
  discovered in the Subaru Deep Field (SDF). The duration of the
  transient is $> \sim800$ d in the observer frame, and the maximum
  brightness during observation reached approximately 23 mag in the
  $i'$ and $z'$ bands. The faint host galaxy is clearly identified in
  all 5 optical bands of the deep SDF images. The photometric redshift
  of the host yields $z\sim0.6$ and the corresponding absolute
  magnitude at maximum is $\sim-20$. This implies that this event
  shone with an absolute magnitude brighter than $-19$ mag for
  approximately 300 d in the rest frame, which is significantly longer
  than a typical supernova and ultra-luminous supernova. The total
  radiated energy during our observation was $1\times10^{51}$ erg.
  The light curves and color evolution are marginally consistent with
  some of luminous IIn supernova. We suggest that the transient may be
  a unique and peculiar supernova at intermediate redshift.

\end{abstract}
\keywords{supernovae:general}

\section{Introduction}

Time-domain surveys in various wavelengths have been making mysterious
new transients discoveries. These results are remarkable, and newly
discovered transients are revolutionizing our knowledge of astronomy
and astrophysics.  The hard X-ray survey of {\it Swift} and related
multi wavelength follow-ups found one unusual transient to be the
tidal disruption of a star by a dormant super massive black hole
\citep{bloom, burrows, levan, zauderer}. This discovery confirmed
tidal disruption events as actual stellar phenomena.  Other candidates
for tidal disruption flares have been reported by optical imaging
surveys \citep{suvi, ps1-10jh, crts-candi, sdss, ptf-candi}.  Optical
untargeted imaging surveys have also been discovering new stellar
explosions.  The discovery of SN2007bi has provided possibly the first
evidence of pair-instability supernova (SNe), which are thought to be
triggered by very massive stars \citep{sn2007bi}.  Ultra-luminous SNe
are another recent discoveries \citep{scp, robert, sn2010gx}. These
transients are characterized by high optical luminosities reaching
peak absolute magnitudes of $-21$ to $-23$.  Because of their
luminosity, they could possibly be detected with an 8-m class
telescope even at higher redshift such as $z\sim4$ \citep{robert,
  tanaka}.

Several other optical transients are theoretically predicted but have
not yet been observationally confirmed. One of them is an orphan GRB
afterglow that is thought to arise as a natural consequence of GRB
jets \citep{rhoads, totani}. However, because of the limited
sensitivity of current optical equipment, there is no promising
candidate of orphan GRB yet.  To extend the redshift frontiers of
these transients and to search for transients undetected by optical
surveys, we performed systematic transient searches and classification
using a deep survey conducted by Subaru/Suprime-Cam. In this letter,
we report the discovery of an unusually luminous long-duration optical
transient. Throughout this paper, magnitudes are in the AB system.

\section{Subaru Deep Field Observations}

We obtained available optical imaging data sets of the Subaru Deep
Field (SDF; \citet{sdf,sdf2}). The original SDF survey was conducted from
2002 to 2003 with 5 broad-band filters--$B$, $V$, $R$, $i'$, and
$z'$-- and two narrow-band filters, NB816 and NB921 using the
Suprime-Cam attached to the Subaru telescope.
The 3 $\sigma$ limiting magnitudes of the final
stacked images reach $B=$28.45, $V=$27.74, $R=$27.80, $i'$=27.43, and
$z'$=26.62, respectively \citep{sdf}.  
The field was also monitored using the same camera before and after
the SDF survey for various purpose with several programs. The total
temporal coverage is $\sim2630$ d from 2001 to 2008.
Although the observations for $B$ and $V$ were made by a single-epoch
observation, multi-epoch data are available for $R$, $i'$, and $z'$
bands. These were particularly suitable for the transient survey.
This field was also observed in U-band by the KPNO Mayall 4 meter
telescope \citep{sdf2}, near-infrared by UKIRT \citep{ukirt,motohara}
and ultraviolet by GALEX \citep{galex}. These multi-wavelength data
sets were crucial for constraining the SED of the transients and/or
their host galaxies.

\section{Analysis and Results}

\begin{figure}
\epsscale{.80}
\plotone{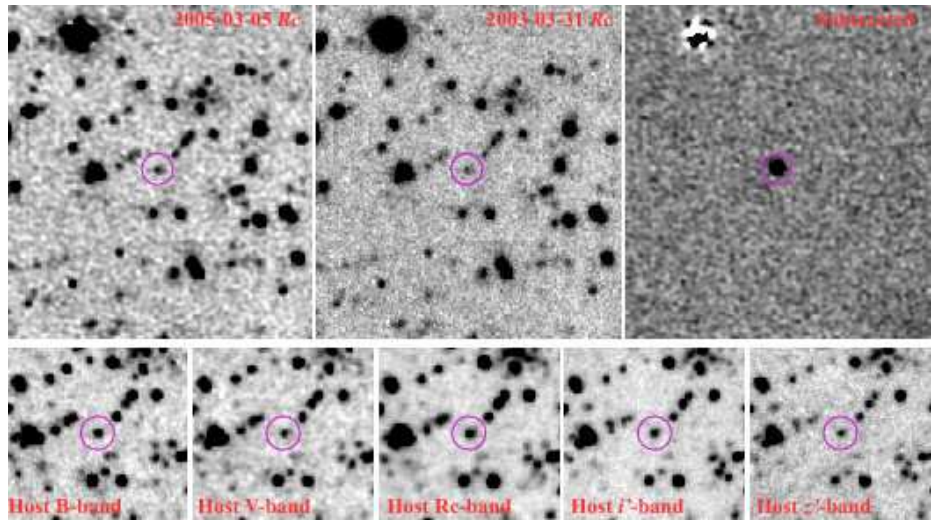}
\caption{The upper panel indicates the discovery of SDF-05M05. The
  left image shows the first detection of SDF-05M05 on 5 March
  2005. Using the image taken on 31 March 2003 (center), we generated
  the PSF-matched subtracted image (right). The subtracted image shows
  the SDF-05M05 clearly. The bottom panel shows the host galaxy of
  SDF-05M05 in the $B$, $V$, $Rc$ $i'$, and $z'$ bands. These images were
  generated excluding the transient component.\label{image}}
\end{figure}

The basic reduction of the Suprime-Cam data was performed using the
SDFRED \citep{ouchi}. To discover the transients and variable
objects, we made nightly stacked images for $Rc$, $i'$, and $z'$.  
%
%
For these 3 bands, differential images were also generated with
special-purpose software tuned to the Subaru/Suprim-Cam data and based
on an algorithm in \citet{diff}.
Figure \ref{image} shows the optical transient found in
images taken on 5 March 2005 at RA$=13^{\rm h}23^{\rm m}52^{\rm
  s}.76$ and Dec$=+27^{\circ}43'58".84$.
The SDSS serendipitously-detected this transient and lists it as a
stellar-like object. We investigated available SDSS images taken on 18
January 2005 with $u'$, $g'$, $r'$, $i'$, and $z'$ band filters. As in
the SDF photometry, we measured the magnitudes in the $r'$, $i'$, and
$z'$ bands on these images. There were marginal detections in the $g'$
and $r'$ bands.
This SDSS observation is the first time this object has been reported
in this position.  Hereafter, we define 18 January 2005 as the
starting point of the transient (T$_{0}$).
Figure \ref{lc} shows light curves for this transient with $R$, $i'$
and $z'$ bands. 
There is no significant variability between March 2001 and April
2003. The amplitude of the event was about 3.4 mag and the transient
faded out until June 2008.
Although the light curves have no temporal coverage for the
brightening period, the duration of the transient is approximately 800
d, which is significantly longer than that for typical Ia, Ib/c and
IIp SNe.

The host galaxy of the transient was identified using the deep-stacked
$B$- and $V$-band images generated by \citet{sdf}. We also made
deep-stack images for the $Rc$, $i'$, and $z'$ bands using images
taken before 5 March 2005 to exclude the transient component. As shown
in Figure \ref{image}, the host galaxy of the transient is discernible
in all 5 bands. We also confirmed that the source is significantly
extended, compared to the size of the point spread function. 
By comparing the positions of nearby stars in our reference frame with
the image in which transient was discovered, we are able to align the
images with accuracy of $0".07$ rms. With an observed offset between
the transient and the host galaxy of $\delta R.A. = +0".15$ and
$\delta Dec. = -0".24$, the likelihood of a nuclear origin for the
transient is only 0.03\%.
%
%
%
These deep images also allowed us to measure the brightness of the
host galaxy accurately. Based on the publicly released SDF catalog, we
made a photometric calibration and performed aperture photometry for
the host galaxy using a 2$"$ radius that was the same as that used for
the SDF catalog.  Although the very deep U-band and NUV (175-275 nm)
images are available for this field, there is no counterpart at the
position.  The $3\sigma$ limits are $26.8$ in U-band and $\sim27$ mag
in NUV, respectively.

\begin{figure}
\epsscale{.80}
\plotone{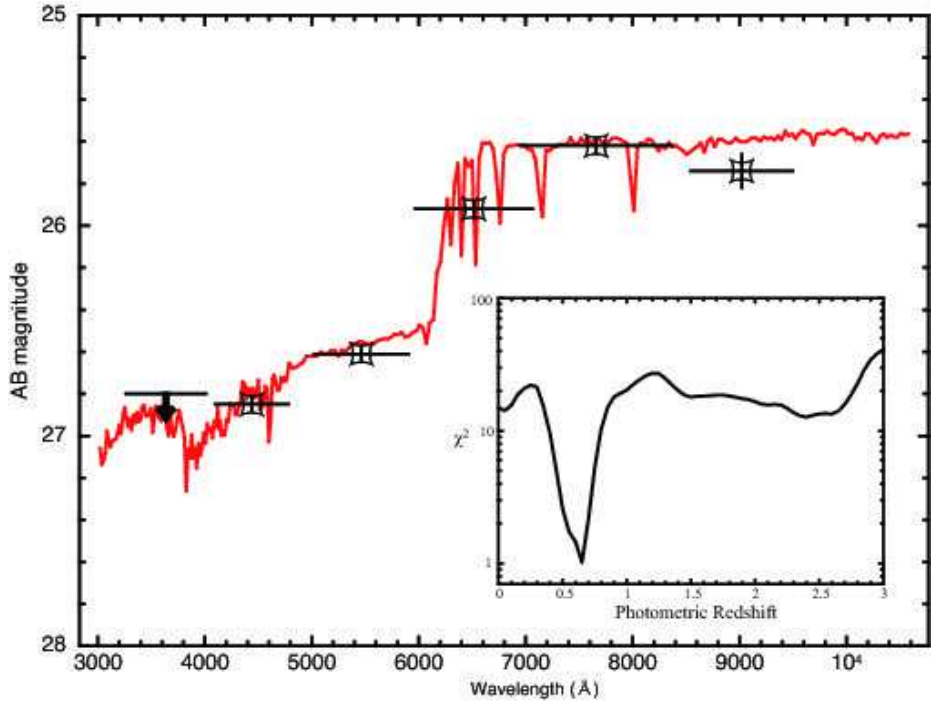}
\caption{The measured $BVRci'z'$ band fluxes of the SDF-05M05 host
  galaxy (data points), compared with the best-fit template SED
  (starburst type) for $z=0.65$ (solid curves). The sub panel shows
  the reduced chi-square of the model fit to the multi-band
  photometric data of the host galaxy, shown as a function of the
  assumed redshift.\label{photz}}
\end{figure}

We estimated a photometric redshift of the host galaxy using Hyper-Z
\citep{hyperz} code. The best-fitting result is
$z=0.65^{+0.02}_{-0.03}$ with reduced $\chi^2$ of 1.02 using the burst
template (age of 0.13 Gyr and the intrinsic extinction $A_{V}$ of 0.2
mag). The 3-sigma error range is 0.50-0.70. Figure \ref{photz} shows
the best-fitting model with actual measurements. 
The age of the host tends to be about 1.5$\sim$2 times older than
those of GRB host galaxies \citep{grbhost}.
We also estimated the photometric redshift by using only the templates
from Coleman, Wu \& Weedman (1980, hereafter CWW). In this case, the
best-fitting galaxy is irregular, and the redshift estimation is
$z=0.62^{+0.02}_{-0.02}$ with a reduced $\chi^2$ of 1.19. In both
cases, the values are in agreement as $z\sim0.6$.  With this
photometric redshift, the peak absolute magnitude of this transient is
estimated to be $M_{R}\sim -20$ mag, which is 2-3 magnitudes brighter
than bright core-collapse SNe. Even at the lower limit of redshift,
the peak absolute magnitude is still $M_{R}\sim -19.3$ mag.
Assuming no bolometric correction, the total integrated optical output
from SDF-05M05 during the observations ($\sim$850 days) was
$\sim1\times10^{51}$ erg, which is comparable with those of GRBs
\citep{grb1, grb2, grb3}.  The absolute host magnitude was also
calculated as $M_{\rm V}= -16.3$ mag, which is comparable to or rather
fainter than that of SMC.

The transient was also imaged in the $J$ and $K$ bands by UKIRT/WFCAM
on 15 April 2005, and the corresponding catalog was generated by
\citep{ukirt}.  At the transient position, we found a point source in
both the J and K bands. Although the contribution of the host galaxy
is unclear in this photometry, the contamination is thought to be
insignificant because the timing of the observation was close to the
peak in the optical light curves (Figure \ref{lc}) and the shapes on
the $J$ and $K$ images are point-like sources while the host in the
optical data is significantly extended.  Furthermore, the expected
magnitudes from the SED fitting of the host galaxy ($J\sim25.5$ and
$K\sim25.2$) are more than 2 mag fainter than the photometric result
of UKIRT. Therefore, the blue of the source could be originating from
the transient.  We also generated the SED of the transient at 88 d
after the first detection with SDF data. As shown in Figure \ref{sed},
the SED is significantly different from power law and well-fitted by a
blackbody with a temperature of $6431\pm310$ K.

\section{Discussions}

We presented detailed SDF data and photometric results of an optical
transient search, which contain evidence of an unusual optical
transient. Its key features are as follows: a long duration light
curve; an intensive absolute magnitude (remaining brighter than $-$19
mag over 300 d); blue SED in NIR data around the first detection; 
  offset from the center of host galaxy; and a faint host
galaxy.
These key features suggest that the unusual transient may be unique
supernova such as an ultra-luminous supernova, or a peculiar supernova
with type IIn spectral features. Below, we discuss differences from
AGN and a tidal diruption flares and possibilities of these two
supernova cases.

\begin{figure}
\epsscale{.80}
\plotone{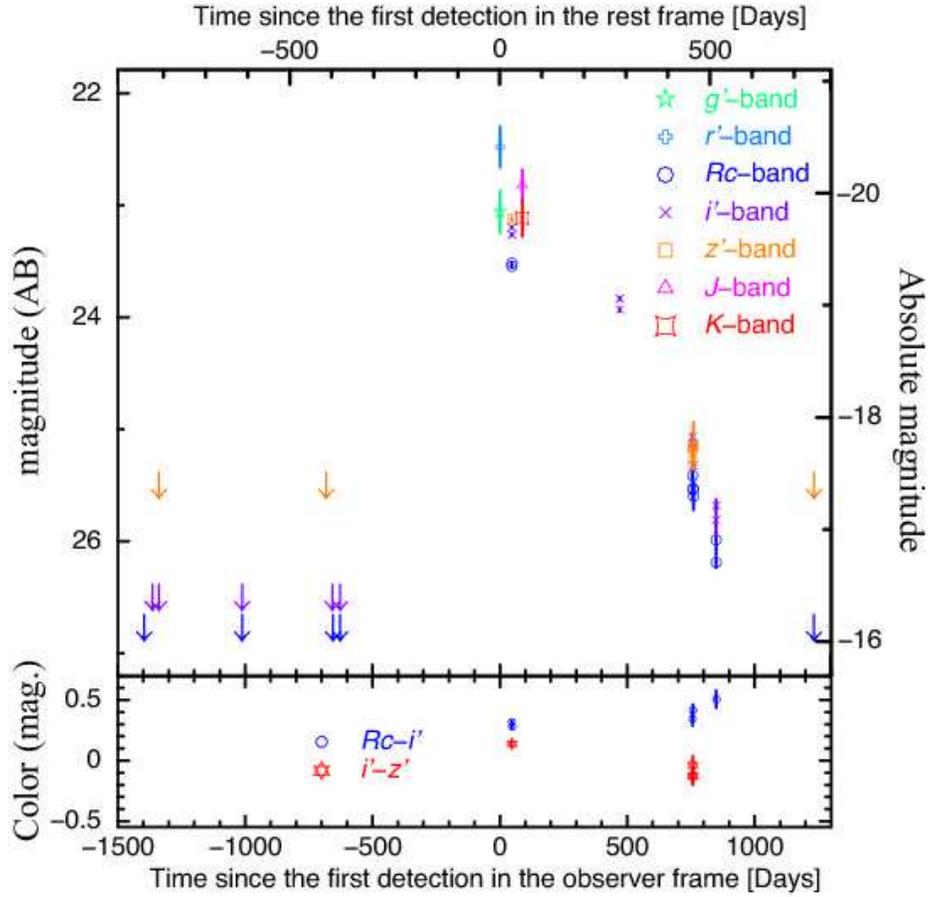}
\caption{Light curves and temporal color evolution of SDF-05M05. SDF-05M05
  maintained a magnitude brighter than $-19$ mag for approximately 300
  d in the rest frame, a period significantly longer than for typical
  SNe or ultra-luminous SNe.\label{lc}}
\end{figure}

AGN origin is unlikely due to the offset from the center of host
galaxy. In addition, the large amplitude of the transient also support
this. Because typical amplidude of AGN variablity is less than 1
mag based on the long term SDSS observations for the Stripe 82 field
(e.g. Ai et al. 2010, Butler \& Bloom 2011, MacLeod et al 2012).
The tidal disruption of stars by massive black holes at the centers of
galaxies show the large amplitude flare at optical, UV, and X-ray
wavelengths although some event showed no UV/transient emission due to a
large amount of dust obscuration \citep{bloom, burrows, ps1-10jh}.
But the tidal disruption flare is also unlikely due to the
offset. Besides the location in the host, the SED and temporal
evolutions are also different from expectation of the tidal diruption
flare at X-ray or UV wavelengths (the temperature of the inner
accretion disc is $\sim3\times10^{5}$ K).  Because the observed
temperature in present case is significantly lower than that predicted
for a tidal disruption, this transient is unlikely to be a typical one
of that type. \citet{tdf-lc} predicted that an early stage
super-Eddington outflow would produce an intensive optical emission
with a blackbody spectrum initially peaking at optical/UV
wavelengths. The expected color evolution becomes bluer if the
observation is made close to the peak wavelength, or showing no colour
change if the observation is on the Rayleigh–Jeans tail.  The
blackbody spectrum peak at around 6500K is consistent with this
prediction \cite{tdf-sed}. However, as shown in Figure \ref{lc}, the
current event shows both bluer and redder colors changing at 750 d
after the first detection. This is inconsistent with the prediction.

\begin{figure}
\epsscale{.80}
\plotone{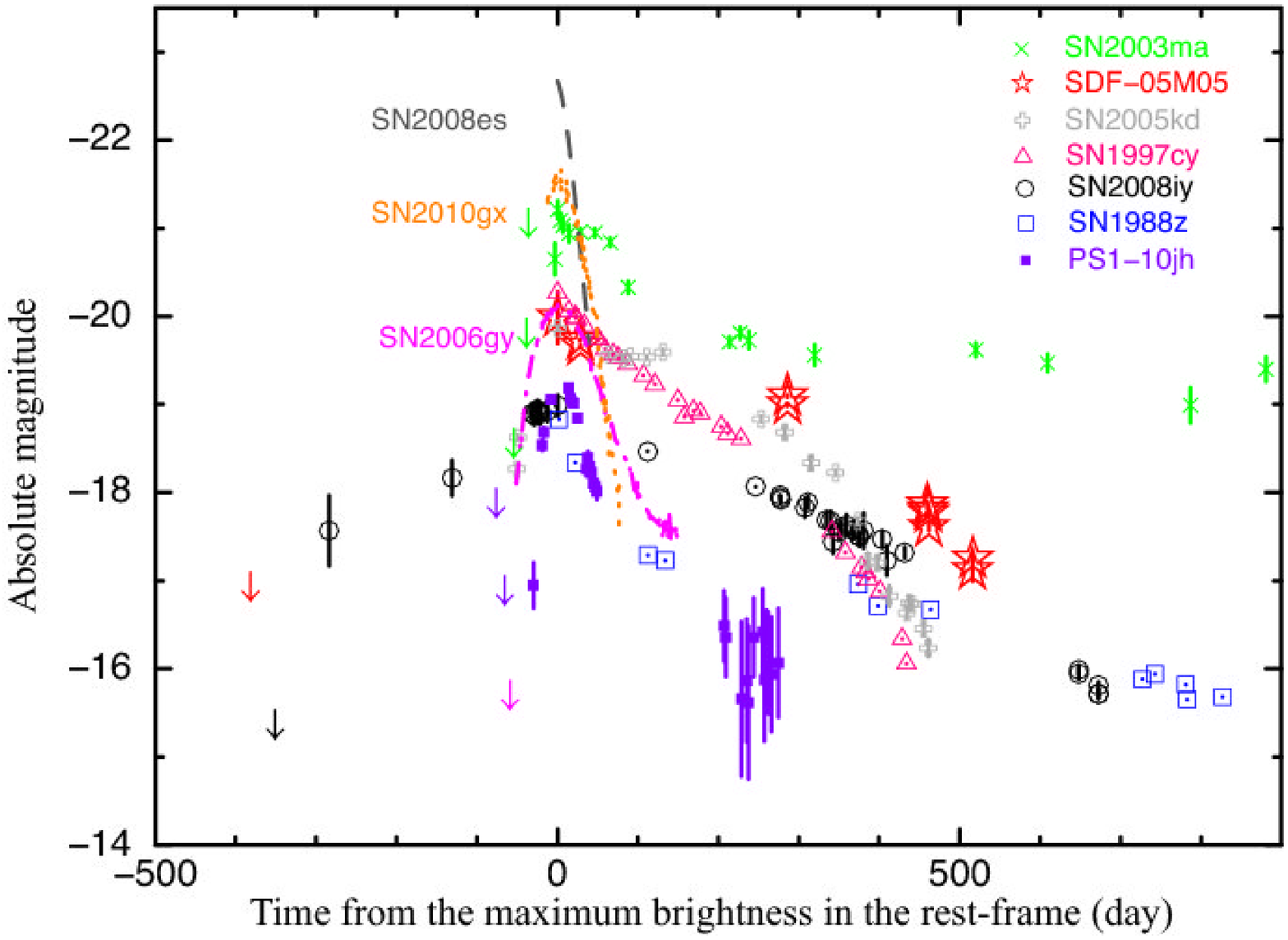}
\caption{Absolute light curve of SDF-05M05 compared to those of tidal
  disruption flare (PS1-10jh; \citet{ps1-10jh}), ultra-luminous
  supernova and peculiar SNIIn. For ultra-luminous supernova, lines
  indicate the light curves of SN2008es \citep{sn2008esa, sn2008esb},
  SN2010gx \citep{sn2010gx}, and SN2006gy \citep{sn2006gy}. For
  peculiar SNIIn, points as labeled show the temporal evolutions of
  SN1988Z \citep{sn1988z}, SN1997cy \citep{sn1997cya,sn1997cyb},
  SN2003ma \citep{sn2003ma}, SN2005kd \citep{sn2005kd}, and SN2008iy
  \citep{sn2008iy}.\label{comp}}
\end{figure}

The key features of ultra-luminous SNe are a roughly symmetric light
curve, an absolute peak magnitudes of $\sim-$21 to $-23$ mag, and a
faint host galaxy (a low mass and presumably a low-metallicity
environment where are desirable for their massive progenitors
(e.g. Stoll et al. 2011; Neill et al. 2011).  Recent systematic
studies by the Palomar Transient Factory, Pan-STARRS and others have
identified a number of such events (e.g., Ofek et al. 2007, Pastorello
et al. 2010 , Quimby et al. 2011; Chomiuk et al. 2011).  To compare
the present case with ultra-luminous SNe, we plotted the light curves
of SN2010gx \citep{sn2010gx}, SCP06F6 \citep{scp}, SN2006gy
\citep{sn2006gy}, and SN2008es \citep{sn2008esa, sn2008esb} together
with that of the current event.
Here, we note that former two events have no obvious evidence of
circumstellar interaction, SN2006gy show it, and SN2008es has
implications.
Figure \ref{comp} shows clear differences in absolute magnitude ($2-3$
mag fainter) and duration ($2-3$ times longer).  The current event is
therefore unlikely to be of ultra-luminous SNe origin.

\begin{figure}
\epsscale{.80}
\plotone{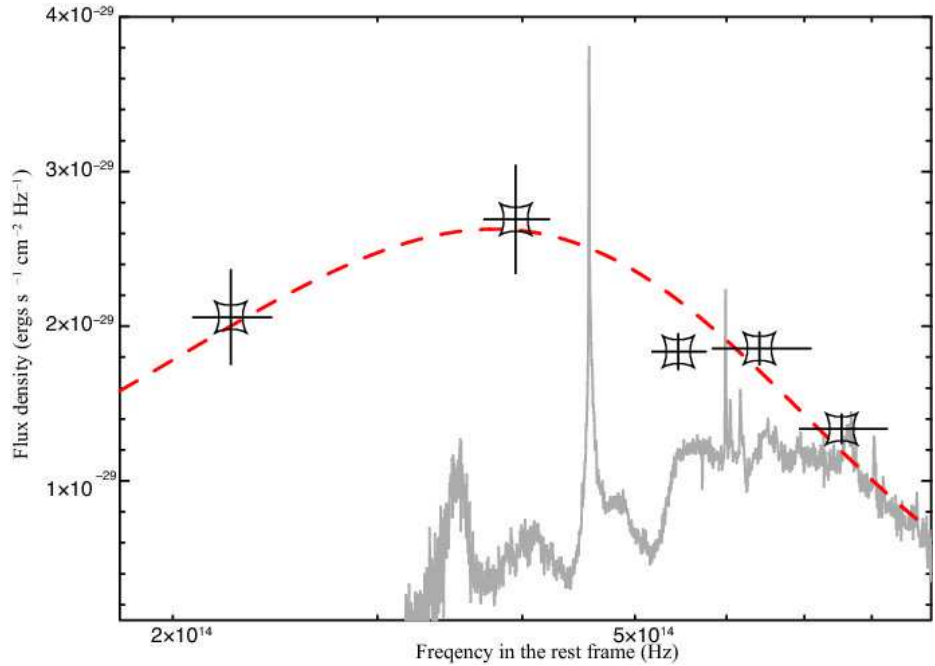}
\caption{The spectral energy distribution of the transient at
  $T_{0}+88$ d. The SED is well fitted by the single-temperature
  blackbody model as described with a dashed line. Box points show the
  photometric results of the SDF-05M05. For the spectral shape
  comparison, solid line show the arbitrarily scaled broadband
  spectrum of SN1997cy at 84 d after the peak.\label{sed}}
\end{figure}

The third possibility is that the observed transient is an SNe with type
IIn spectral features. Such events are still rarely observed, but the
number of detections is increasing.
Figure \ref{comp} shows the light curves of SN1997cy \citep{sn1997cya,
  sn1997cyb}, SN2003ma \citep{sn2003ma}, SN2005kd \citep{sn2005kd},
and SN2008iy \citep{sn2008iy}. All 4 events show extremely long
($>\sim400$ d) durations.
For SN2008iy, additional data points were collected from the
3$\pi$-survey of Pan-STARRS1 \citep{kaiser}. This was because the time
coverage of the data available from literature was insufficiently
long to allow a comparison with the current event.  The survey
successfully detected the late phase (490$\sim$1000 d after the peak)
of SN2008iy in $i_{P1}$ and $z_{P1}$ band \citep{ps1phot}.
Besides SN2003ma, the temporal evolution of the current event
resembles those of SN1997cy, SN2005kd and SN2008iy in the temporal
breaks at around $\sim300-500$ d in the rest frame. The linear decline
rates of SN1997cy and SN2005kd before and after the break of these
events are commonly $\sim0.65$ mag/100 d and $\sim1.5$ mag/100 d,
respectively. The latter is faster than the expected rate of decline
for radioactive $^{56}{\rm Co}$ (0.98 mag/100 d).  Although the
temporal duration of the current event is comparable to those of
peculiar SNe, there are significant differences in the decline rates.
The linear decay rates of SDF-05M05 in the $i'$-band before and after
the temporal break are 0.28 mag/100 d and 0.76/100 d,
respectively. These decline rates are all slower than the expected
rate of the decline for radioactive $^{56}{\rm Co}$. 
SN2008iy has a decline rate similar to that of SDF-05M05. However, the
broad-band SED of the current transient is well fitted by the
single-temperature blackbody model, whereas that of SN2008iy deviates
from the blackbody and is more like that of usual type IIn event
\citep{sn2008iy}.
As shown in Figure \ref{sed}, this broad-band SED property is also
another significant differences between SN type IIn and SDF-05M05.
In addition, the temperature was lower than that of Type IIn SN2003ma
and SN2008am, which had blackbody spectrum \citep{sn2003ma,2008am}.
Although we cannot entirely rule out the possibility of this event
being of peculiar SNe-IIn origin because of unclear common properties
of peculiar-IIn events, the current energetic event may be a new type
of optical transient altogether.

All of these luminous optical transients are found in faint host
galaxies. Although this may be due to selection bias, the
characteristics of the host galaxies are crucial to an understanding
of the origins and occurrence rates of the transients.
Considering to the luminosity function of various types of galaxies
(e.g Zucca et al. 2006), the fraction of faint galaxies, such as the
host galaxy of SDF-05M05, is expected to increase with redshift.
\citet{hostlf} used the same 4 CWW galaxy templates used for the
photometric redshift in the present study.
This may make detections of luminous events rare in nearby Universe but
common at higher redshift (e.g., $z>\sim0.5$). The apparent magnitudes
at higher redshift will be fainter than 22-23 mag as same as
in the current event. 
The limiting magnitudes of the medium deep survey of Pan-STARRS is
comparable with the maximum brightness of the current event. Hence, it
is expected that Pan-STARRS has been detecting numbers of these long
and luminous optical transients associated with faint host galaxies.
However, a slow evolution and the presence of a faint host galaxy make
it difficult to classify these events as real optical
transients. Therefore, coordinated long-term monitoring with
larger-aperture telescopes is needed to better determine their
origins. In such work, the planned strategic survey with a
time-domain-survey cadence using the new wide-field imager,
Hyper-Suprime-Cam attached to Subaru will prove
invaluable.

\acknowledgments

This work is supported by grants NSC 100-2112-M-008-007-MY3 (YU),
99-2112-M-002-002-MY3 (KYH) and the Grant-in-Aid for Scientific
Research (19540238) from the JSPS of Japan.
The United Kingdom Infrared Telescope is operated by the Joint
Astronomy Centre on behalf of the Science and Technology Facilities
Council of the U.K. Use of the UKIRT 3.8-m telescope for the
observations is supported by NAOJ.
The PS1 Surveys have been made possible through contributions of the
Institute for Astronomy, the University of Hawaii, the Pan-STARRS
Project Office, the Max-Planck Society and its participating
institutes, the Max Planck Institute for Astronomy, Heidelberg and the
Max Planck Institute for Extraterrestrial Physics, Garching, The Johns
Hopkins University, Durham University, the University of Edinburgh,
Queen's University Belfast, the Harvard-Smithsonian Center for
Astrophysics, and the Las Cumbres Observatory Global Telescope
Network, Incorporated, the National Central University of Taiwan,
and the National Aeronautics and Space Administration under Grant
No. NNX08AR22G issued through the Planetary Science Division of the
NASA Science Mission Directorate.

\end{document}